\documentclass[10pt,conference]{IEEEtran} 
\usepackage[letterpaper,left=0.625in, right=0.7in, bottom=1in, top=0.7in]{geometry}

\usepackage{graphicx} % Required for inserting images
\usepackage{hyperref}
\usepackage[cmex10]{amsmath}
\usepackage{%
	amsfonts,%
	amsmath,%
	amssymb,%
	amsthm,%
	bbm,%
	cite,%
	float,%
	etoolbox,%
	mathrsfs,%
	mathtools,%
	multirow,%
	pgf,%
	pgfplots,%
	subfigure,%
	tikz,%
}
\usepackage{algorithm}
\usepackage{algpseudocode}
\usepackage{bm}
\usepackage{tikz}
\usepackage{amssymb}

\usepackage{graphicx}
\usepackage{subfigure}
\usepackage{subcaption}
\newtheorem{rem}{Remark}

\newtheorem{theorem}{Theorem}%[section]

\title{
The Proportional Fair Scheduler in Wavelength-Multiplexed Quantum Networks
}

\author{
    \IEEEauthorblockN{Sanidhay Bhambay\IEEEauthorrefmark{1}, Siddarth Koduru Joshi\IEEEauthorrefmark{2}, Thirupathaiah Vasantam\IEEEauthorrefmark{3}, Neil Walton\IEEEauthorrefmark{1}}
    \IEEEauthorblockA{\IEEEauthorrefmark{1}Durham University Business School}
        \IEEEauthorblockA{\IEEEauthorrefmark{2}Quantum Engineering Technology Labs, School of Electrical, Electronic and Mechanical Engineering, University of Bristol}
    \IEEEauthorblockA{\IEEEauthorrefmark{3}Department of Computer Science, Durham University}

}

\begin{document}
% Wavelenth multplex
\maketitle
\begin{abstract}
We address the problem of optimal pumping strategies in quantum networks. These networks enable secure communication by distributing entangled photon pairs to user (or node) pairs. Quantum Key Distribution (QKD) protocols, like BBM92, generate secret keys from entangled photons. While secure communication and error correction are essential for any quantum communication channel, resource contention, optimization, and fairness issues are critical for networks.  
In this article, we analyze the performance of quantum networks, proposing simple distributed algorithms for QKD networks generating secret keys.

There are significant advantages of pumping entangled photons in QKD networks, but challenges arise in practical implementations. The underlying channels are inherently time-varying, and thus data rates fluctuate between nodes.
Moreover, multiple edges (node pairs) can be pumped simultaneously, albeit at the cost of a reduced secret key rate (SKR). These temporal and spatial constraints yield a complex decision-making problem whose solutions may favor a small set of user pairs to the detriment of overall, long-run network performance.

We design adaptive pumping strategies that address these challenges in QKD networks. 
In particular, we find that a proportional fairness pumping strategy (PF-PS) stands out by dynamically prioritizing users with lower average secret key rates and optimally balancing fairness with throughput. The proposed algorithm is a natural extension to quantum networks of the Proportional Fair Scheduler deployed in 4G LTE and 5G mobile networks.
Both theoretical analysis and numerical simulations confirm that PF-PS is optimal for entangled state distribution, and thus, when adapted appropriately, proportional fair pumping is a strong candidate for efficient resource allocation in quantum networks.
% It is especially effective for QKD applications.

% % Setting 
% We address the problem of optimal switching strategies in quantum networks (QNs). A quantum network distributes entangled states for tasks such as Quantum Key Distribution (QKD). QKD protocols, such as BBM92,  require bi-partite entanglement distribution, and thus efficient distribution of entangled photons between user pairs. 

% % Issues
% Network conditions are time-varying, and 
% The network is a time-varying graph, where edges represent active connections between users formed by the distribution of bi-partite entangled states between the pair of users. We solve an optimization problem that involves selecting connections between user pairs at each time step to maximize a given utility function. 

% % Contribution and methods
% Specifically, we considered three switching strategies for connecting different user pairs namely: Proportional Fairness (PF), Max-Min, Greedy, and Round-Robin. The PF scheme dynamically prioritizes user pairs with low historical average Secret Key Rates (SKRs), achieving a balance between fairness and throughput. Both theoretical analysis and numerical results demonstrate that the PF scheme is more beneficial for generating keys in QCN with QKD nodes. 

\end{abstract}

% \neil{QKD makes sense, but how common is the abbreviation QN for Quantum Network? The term quantum network seems quite generic, but when we use the abbreviation QN, it feels quite specific. PErhaps we should just say Quantum network. Or maybe use QLAN for Quantum Local Area Network.}

\section{Introduction}
% QKD~\cite{gisin2002quantum,scarani2009security} has emerged as a main backbone of quantum-secure communications, with significant advancements in both research and real-world implementations. In Europe, the INT-UQKD project a collaborative initiative led by ESA and partners like SpeQtral and evolutionQ aims to establish a hybrid quantum network integrating terrestrial fiber and satellite-based QKD. This network connects trusted nodes in Luxembourg, Belgium, and Singapore, with plans for global expansion, demonstrating the feasibility of quantum-secure communications across continents~\cite{INTUQKD2025}. Similarly, the UK Quantum Network (UKQN) has develop fiber-based QKD networks in Bristol and Cambridge, leveraging the National Dark Fibre Facility (NDFF) to achieve SKRs exceeding 1 Mb/s over 410 km. These networks integrate QKD with classical traffic, validating scalability and interoperability with existing infrastructure~\cite{FibreBasedQKD,QCHubWP1,QuantumNetworksUK}. 
QKD has emerged as a main backbone of quantum-secure communications, with significant advancements in both research and real-world implementations~\cite{gisin2002quantum,scarani2009security}. Several projects have showcased the current state of QKD implementations. For example, the INT-UQKD project, a collaborative European initiative led by ESA with partners such as SpeQtral and evolutionQ, aims to establish a hybrid quantum network (QN) that integrates terrestrial fiber and satellite-based QKD across trusted nodes in Luxembourg, Belgium, and Singapore, with plans for global expansion~\cite{INTUQKD2025}. Similarly, the UK Quantum Network (UKQN) has developed fiber-based QKD networks in Bristol and Cambridge, leveraging the National Dark Fibre Facility (NDFF) to achieve secret key rates exceeding 1 Mb/s over 410 km while integrating QKD with classical traffic~\cite{FibreBasedQKD,QCHubWP1,QuantumNetworksUK}. Although these trusted-node architectures demonstrate significant progress, they inherently compromise end-to-end security by introducing potential vulnerabilities at intermediary points. In contrast, entanglement-based QKD protocols eliminate the need for trusted nodes by leveraging the intrinsic security of quantum entanglement, thereby paving the way for more robust and scalable QNs.

% \s{Why are we focusing on trusted node QKD when we are specifically crafting our protocols for entanglement based QKD, i.e. without trusted nodes? We should just say something like Trusted node stuff is a security compromise and better networks exist. Unless you are also targeting trusted node stuff as well. In which case we need to change much of the text.}

QKD enables secure key exchange by exploiting the quantum properties of entangled photons. It achieves information-theoretic security without relying on classical cryptographic systems~\cite{scarani2009security,diamanti2016practical}. A QKD protocol generates quantum states encoded in a photon (in mutually orthogonal bases such as polarization, phase, or time bin). These photons are then transmitted through a quantum channel. At the receiver’s end, these photons are measured in one of a set of randomly selected and orthogonal bases, a process that introduces uncertainty to any eavesdropper. Following this quantum exchange, the communicating parties engage in classical communication and post-processing steps such as sifting, error correction~\cite{shor1996fault}, and privacy amplification~\cite{renner2008security} to generate a mutually shared symmetric and information theoretically secure string of bits called the secret key.

The fundamental limit in any QKD network is imposed by the loss in the optical channel. It is known that for a link with transmissivity $\eta$, the two parties cannot generate more than $-\log(1-\eta)$ secret bits per quantum channel usage~\cite{pirandola2017fundamental}. 
To overcome this limitation, one can use quantum repeaters~\cite{briegel1998quantum,dur1999quantum,van2014quantum} 
% to construct quantum networks~\cite{van2014quantum} that can be used for the distribution of quantum keys.\textcolor{blue}{This argument should be clarified, not clear enough now} These technologies aim to divide a long lossy channel into shorter segments where loss is less severe, enabling techniques like entanglement swapping and purification to bridge the gaps and effectively boost the SKR over long distances.
% \s{Again I am not sure wht the point of discussing routers etc is.}

Another important challenge in practical QKD implementations, the raw keys generated are of finite length, which introduces uncertainties that do not appear in the asymptotic (infinite-key) regime assumed by many theoretical security proofs~\cite{lo2005efficient}. When few qubits have been exchanged, the estimation of parameters such as the Quantum Bit Error Rate (QBER) becomes less accurate. Therefore, as the QBER increases, either due to channel noise or active eavesdropping, more bits must be sacrificed during error correction and privacy amplification to ensure that the final key is secure. To mitigate finite-key effects, a strategy is to accumulate larger sifted key batches before processing. This improves parameter estimation accuracy and optimizes error correction/privacy amplification, thereby increasing the net SKR~\cite{scarani2009security}.

% Consequently, a higher QBER reduces the fraction of bits that can be securely retained. For example in the specific case of the BBM92 protocol, security proofs indicate that if the QBER exceeds approximately 11\%, the net SKR drops to zero~\cite{}.
% \s{Could we perhaps discuss the specific case of BBM92. But the 11\% theory limit does not have anything to do with the finite key effect. The point of the finite key effect is that the processing efficiency of the key depends strongly and non linearly on the length of the the sifted key. Thus, waiting for a long time before utilizing a batch of quantum measurements is a good strategy. But, once information has been converted from quanutm to classical then it is vulnerable to all cyber security risks. Thus for practical concerns the age of the key is important. Hence there is a trade-off between SKR and age of the key. }
% Unlike BB84 protocols there are entanglement-based QKD protocols such as the E91 protocol introduced by Eker~\cite{ekert1991quantum}. The E91 protocol utilizes entangled photon pairs and Bell's inequality to ensure that any interception attempts by an eavesdropper are detectable anomalies. 

Significant efforts have focused on improving SKRs in entanglement-based QKD (like BBM92) by addressing finite-key effects and optical fiber losses. However, while advancements in protocol security and key rate optimization have been substantial, the challenge of ensuring \textit{fairness} equitable key distribution among multiple users remains underexplored in QNs. Fairness, a well-established concept in classical network resource allocation~\cite{kelly1997charging}, ensures that all participants receive adequate resources despite heterogeneous channel conditions. In classical contexts, such as cellular networks, fairness mechanisms dynamically allocate bandwidth to balance throughput and user rates~\cite{viswanath2002opportunistic,andrews2001providing}. Recent works have begun to investigate fairness in QNs~\cite{maule2024fair,vardoyan2023quantum}, yet a comprehensive approach for dynamic resource allocation, and equitable key distribution remains an open challenge.

To address fairness challenges, we consider a QN in which a single-photon source dynamically distributes entangled pairs among $n$ QKD nodes (see Figure~\ref{fig:QCN}). For each pair of nodes $(i,j)$, the SKR is determined by the product of the raw key rate and a reduction factor that accounts for practical imperfections such as noise and potential eavesdropping, as captured by the QBER and its associated binary entropy function. In other words, while the raw key rate benefits from higher channel transmissivity, any increase in errors (or QBER) effectively reduces the overall key rate.
Moreover, the channel conditions including QBER, and other factors vary over time, leading to stochastic fluctuations in the instantaneous SKR. This time-varying nature of the channels further complicates the task of achieving both high SKR and fairness.

An important question arises \textit{how can we maximize the overall SKR across all QKD node pairs while ensuring fairness, despite these time-varying channel conditions?} Since the photon source is capable of pumping entangled photons to multiple pairs of nodes simultaneously, the challenge lies in optimally scheduling the entangled-pair distribution. Specifically, we must determine which combinations of node pairs should be served at each time instance taking into account the current channel state to maximize the secret key generation efficiency while ensuring a fair allocation of resources among all users.

For distributing entangled photon pairs, the source uses wavelength multiplexing~\cite{wengerowsky2018entanglement} to produce a hyper-entangled state and route each photon to a different wavelength channel, forming a single link per distribution. Unlike classical networks where additional links boost capacity, QKD networks are limited by the number of single-photon detectors. As shown in \cite{wang2022dynamic}, adding extra links increases noise (QBER) due to detector timing limitations, which can negate any potential key rate gains. Therefore, to optimize the balance between signal and noise, we consider only a single entangled link per QKD node pair.

To address the dual challenges of maximizing SKR and ensuring fairness in entangled pair distribution, we propose a \textit{proportional fairness pumping strategy} (PF-PS) that dynamically schedules photon pair allocations based on both instantaneous and average key rates while explicitly accounting for time-varying channel conditions. Inspired by classical network fairness frameworks~\cite{kelly1997charging}, this strategy balances throughput and fairness by prioritizing QKD node pairs with high instantaneous SKR relative to their historical averages. Under the PF-PS, the optimal topology for pumping entangled photons is selected using
\begin{equation}
\label{eqn:pf_g}
    G^\star = \arg\max_{G \in \mathcal{G}} \sum_{\substack{i,j=1 \\ i\leq j}}^{n} \frac{S_{ij}(G,\bm\chi)}{\bar{x}_{ij}},
\end{equation}
where $\bm \chi$ denotes the current channel state, $\mathcal{G}$ denotes the set of feasible topologies supported by the photon source and $\bar{x}_{ij}$ represents the average SKR for the node pair $(i,j)$. 
In this formulation, the instantaneous SKR depends on both the network topology $G$ and the channel conditions $\bm\chi$, as these factors jointly determine the pumping of entangled photons. PF-PS inherently penalizes under-allocated links (low $\bar{x}_{ij}$) while rewarding those with high instantaneous SKR $S_{ij}(G,\bm \chi)$, dynamically adjusting allocations to favor fairness without sacrificing overall efficiency. This approach ensures that all users receive equitable key distribution over time, even under heterogeneous and fluctuating channel conditions a critical advancement for scalable, multi-user QKD networks.
% Note that in the above expression, the instantaneous SKR depends on $G$ because the network topology $G$ determines which node pairs receive entangled photons from the source.
% PF-PS inherently penalizes under-allocated links (low $\bar{x}_{ij}$) while rewarding those with high instantaneous SKR $S_{ij}$, dynamically adjusting allocations to favor fairness without sacrificing global efficiency. This approach ensures that all users receive equitable key distribution over time, even under heterogeneous channel conditions a critical advancement for scalable, multi-user QKD networks.

Despite PF-PS theoretical promise, practical QKD networks face constraints like the photon source’s finite capacity and heterogeneous channel conditions inherently create disparities in SKRs across nodes. Existing approaches often prioritize throughput at the expense of fairness, leaving some node pairs underserved. To systematically address this imbalance, we rigorously analyze PF-PS against alternative strategies, establish its fairness and validate its effectiveness across dynamic network conditions.

\subsection{Our Contributions}
In this work, we consider the problem of fairness in a QN where each pair of QKD nodes strives to generate secret keys under time-varying channel conditions. In practical networks, the photon source has limited capacity and stochastic SKR fluctuations.
% , and the optical channels exhibit varying losses and error rates. 
This situation leads to disparities among users, with some experiencing significantly lower SKRs than others. To address this problem, our contributions are as follows:

\begin{enumerate} 
\item We formulate an optimization problem for pumping entangled photon pairs to QKD nodes explicitly accounting for time-varying channel. Specifically, we determine which QKD node pairs should be allocated entangled photons from a single source to maximize a utility function that captures the SKRs of each pair while also ensuring fairness in distribution of limited resources. 
% This optimization is subject to the capacity constraints of the photon source, which can serve only a limited number of node pairs simultaneously, thereby balancing overall performance with equitable access.
\item We propose a gradient-based pumping algorithm that iteratively converges to the solution of the utility-based optimization problem. To the best of our knowledge, this is the first work to introduce a gradient-based algorithm for utility optimization in a QN of QKD nodes. Our method adapts to real-time channel variations, solving the fairness-throughput trade-off without requiring prior knowledge of channel statistics.
\item By choosing different utility functions within our gradient-based framework, we derive and propose several pumping strategies. These include the Proportional Fairness Pumping Strategy (PF-PS), Greedy Pumping Strategy (G-PS), and Round-Robin Pumping Strategy (RR-PS), each designed to address different aspects of the fairness-throughput trade-off.
% \item We investigate several pumping strategies namely Proportional Fairness Pumping Strategy (PF-PS), Greedy Pumping Strategy (G-PS), and Round-Robin Pumping Strategy (RR-PS) that consider both fairness and overall throughput (total SKR) across the network. 
\item We prove that, under the PF-PS, the QN maximizes the geometric mean of the SKRs among all node pairs, thereby ensuring a balanced distribution of resources and fairness. 
% \item To the best of our knowledge, this is the first work to propose a gradient-based algorithm for utility optimization in QN. Our method adapts to real-time channel variations, solving the fairness-throughput trade-off without prior knowledge of channel statistics.  
% \item We support our theoretical findings with extensive simulations. In our study, we consider two scenarios. One in which the channel conditions are fixed and another in which they vary due to changes in the QBER. In both cases, we demonstrate that the PF-PS outperforms the other pumping strategies, validating its effectiveness in balancing throughput and fairness.
\item Extensive simulation results confirm that PF-PS consistently outperforms alternative strategies in balancing fairness and throughput, under both fixed and dynamically varying channel scenarios.

\end{enumerate}

\subsection{Related Works}
% The advancement in the QKD and the quantum networks has been shaped by foundational theoretical work, experiments and practical deployments. In~\cite{bennett1984update}, the BB84 protocol was introduced, leveraging the quantum properties of single photons to enable information-theoretically secure key exchange over untrusted channels. Subsequently, in~\cite{ekert1991quantum} proposed entanglement-based QKD (E91), where correlated measurements of entangled photon pairs ensure security. 

Existing literature largely treats QKD networks as either static topologies~\cite{bennett1984update,ekert1991quantum,scarani2009security} or isolated point-to-point links, often neglecting the interplay between dynamic channel conditions (e.g., time varying QBER) and resource constraints. Recent works, however, have begun addressing utility maximization in quantum networks. For instance,~\cite{vardoyan2023quantum} extends the classical concept of Network Utility Maximization (NUM) to quantum networks by introducing three entanglement-based utility functions: distillable entanglement, secret key fraction, and entanglement negativity. In this approach, the NUM problem is formulated by assessing the utility of individual routes based on the rate and fidelity of the allocated entanglement. The study shows that, depending on the chosen entanglement measure, the NUM problem may not be convex. Furthermore,~\cite{kar2024convexification} demonstrates that in static scenarios where routes and applications are fixed, meaning that a route's utility changes only when its resource allocation is modified the formulation becomes convex.
% thereby analyzing trade-offs between rate and fidelity in resource allocation; while closely related to our work, this study focuses on entanglement metrics rather than the SKR dynamics prioritized in our model
In~\cite{lee2024quantum} authors introduces a utility-based framework to quantify quantum network value across applications such as secure communications and distributed quantum computing.
In this work a route’s utility is mainly determined by the execution rate of its corresponding task and the associated computational demands. Consequently, the network utility is defined as the highest total of route utilities achievable under feasible rate allocations.
% , with scaling laws suggesting that distributed quantum computing offers superior utility potential compared to classical systems. 
In~\cite{maule2024fair} authors proposes fairness-aware scheduling strategies for satellite constellations to optimize secret key generation among ground stations, although their approach relies on empirical scheduling without explicitly modeling SKR dynamics (e.g., QBER) or employing utility-driven optimization. Finally, ~\cite{pouryousef2024resource} presents a network planning framework for optimizing quantum repeater placement in existing infrastructure by evaluating the impacts of memory multiplexing, coherence time, and fairness assumptions on entanglement distribution utility.
\subsection{Organization}
The rest of the paper is organized as follows. In Section~\ref{sec:model}, we describe the model under consideration and describe the system state descriptor. We formulate the optimization problem for pumping entangled photons in Section~\ref{sec:problem_statement}. Furthermore, we describe the gradient based pumping algorithm to the solve the optimization problem in Section~\ref{sec:gradient_algo}. In Section~\ref{sec:schemes}, we discuss different pumping strategies used in this work. Our theoretical findings on the PF-PS are presented in Section~\ref{sec:th_results}. In Section~\ref{sec:numerical_studies}, we provide numerical results to support our claims. Finally, we conclude our work in Section~\ref{sec:conclusion} and outline future directions.
\subsection{Notations}
We denote $[M]$ to represent $\{1,\ldots, M\}$. 
% We use $\bar{\R}$ to denote the set of extended reals, i.e., $\R\cup \{\infty\}$. For $x,y \in \mathbb R$, $x\vee y = \max\{x,y\}$ and $x\wedge y=\min\{x,y \}$. For $i \in [M]$, $\bm e_i$ is the $i$-th unit vector in $\mathbb R^M$; $\bm 1$ is the vector of all ones; $\bm 0$ is the zero vector. 
For vectors $\bm u, \bm v \in \mathbb R^M $, we write $\bm u \leq \bm v$ if $u_i \leq v_i$ for $i\in[M]$. For a set $\mathcal{M}$, we use $|\mathcal{M}|$ to denote its cardinality. We use $\bm a \cdot \bm b$ to denote the dot product and $\bm a \bm b$ to represent the element-wise multiplication between two vectors of the same dimension. For a function $f(x)$ we use $f'(x)$ to denote its derivative with respect to $x$. 
% We use the notation $[y]_+=\max(y,0)$ and $[y]_+^a=\max(y,a)$.

\section{System model and state description}
\label{sec:model}
\begin{figure}[h!]
  \centering
  \includegraphics[width=9cm]{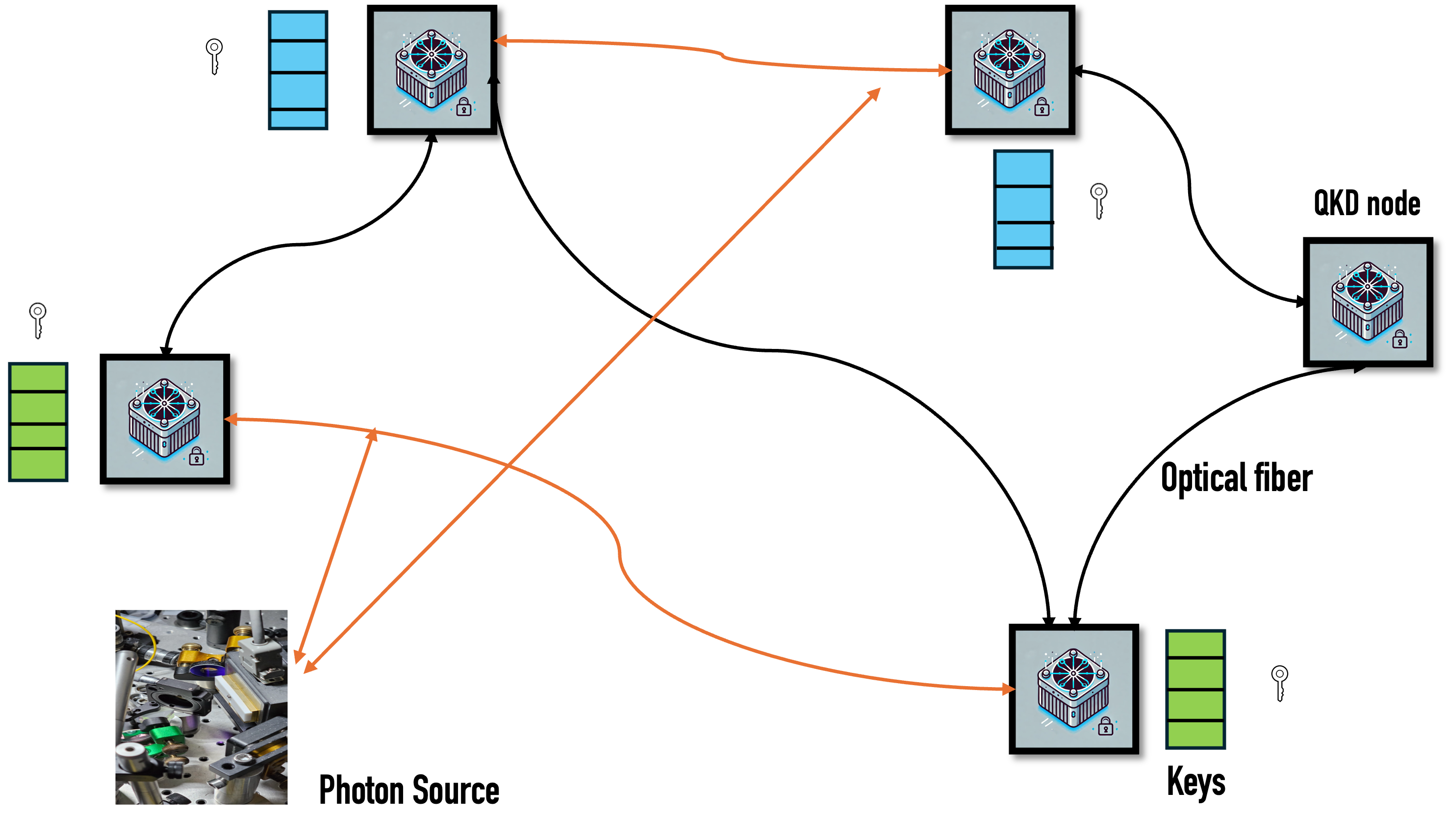}
\caption{A quantum network consisting of QKD nodes and a single polarization-entangled photon source is depicted. The orange edges connecting pairs of QKD nodes indicate that the source is delivering entangled photons to those node pairs.}
\label{fig:QCN}
\end{figure}

In this section, we describe a QN that pumps entangled photons to pair of QKD nodes, as shown in Fig.~\ref{fig:QCN}. First, we present the system model considered in this work. Specifically, we discuss the entangled-photon source used to pump pairs of entangled photons, the QKD nodes, and the processes involved in generating keys between pairs of QKD nodes. We then introduce the state descriptor used to represent our QN. Finally, we describe the SKR rate region. 
% \neil{discuss organization of this section.}

% \textcolor{red}{Why there are two different colors for edges in the figure? What is the significance?}
\subsection{System Model}
Each QKD node in the QN aims to establish secure communication with every other node. This is achieved by generating secret keys using BBM92 protocol between different pairs of nodes.

To facilitate key generation using the BBM92 protocol between nodes, bipartite entanglement states must be shared between them. This is accomplished by distributing entangled photons from a polarization-entangled photon source, as shown in Fig~\ref{fig:QCN}. The source generates entangled photon pairs, with one photon sent to each of the two nodes involved in the communication.
For example, a commonly used entangled state of photon pairs is
$$
|\psi^{+}\rangle = \frac{1}{\sqrt{2}} \left( |HH\rangle + |VV\rangle \right),
$$
where $|H\rangle$ and $ |V\rangle $ represent the horizontal and vertical polarization states of the photons.
The distributed keys are subsequently used to encrypt and decrypt messages, enabling secure communication between the users. Formally, we consider a QN with
$n$ QKD nodes and a single polarization-entangled photon source.

The QN considered in this work has three important components, described below.
% These secret keys are then used for encryption and decryption, ensuring secure communication between users. 
% The efficiency of key generation is characterized by the SKR, defined as the number of secret key bits generated per second between a pair of nodes.
% We denote the instantaneous SKR between a pair of users $i$ and $j$ by $ \text{SKR}_{i,j}$.
\subsubsection{Entangled-Photon-Source} The polarization-entangled photon source generates entangled photon pairs via spontaneous parametric down-conversion (SPDC) in a nonlinear crystal. A pump laser illuminates the crystal, causing a probabilistic conversion of high-energy photons into pairs of lower-energy photons with correlated polarization states. For the Bell state $|\psi^{+}\rangle = \frac{1}{\sqrt{2}} \left( |HH\rangle + |VV\rangle \right)$, the crystal is configured to produce horizontally $H$ and vertically $V$ polarized photon pairs with equal probability.

In this work, we assume that the photon source has limited capacity and can pump entangled photons to at most $C$ pairs of QKD nodes at a time.
\subsubsection{QKD Node}
Each QKD node consists of a beam splitter with $H/V$ or $+/-$ basis selection and single-photon detectors. Further, each QKD node has a classical channel for post-processing (sifting, error correction, privacy amplification).

Nodes are interconnected via optical fibers, with channel transmissivity $\eta_{ij} = 10^{-\beta L_{ij}/10}$, where $\beta$ (dB/km) is the fiber attenuation coefficient and $L_{ij}$ is the distance between nodes $(i,j)$.  
\subsubsection{Secrete Key Generation}
The process of generating a secret key between different pairs of QKD nodes follow the steps listed below:
\begin{itemize}
\item Raw Key Extraction:
Nodes measure entangled photons using randomly chosen bases. After reconciling their measurement choices, they keep only the results where their bases match, forming a raw key.
\item Error Correction:
A portion of the raw key is publicly compared to estimate errors. Errors are then corrected to ensure both nodes have identical keys while minimizing information leakage to potential eavesdroppers.
\item Privacy Amplification:
To eliminate any remaining knowledge an eavesdropper might have gained, the key is further processed to shorten it while ensuring strong secrecy guarantees.
\end{itemize}
The efficiency of key generation is characterized by the SKR, defined as the number of secret key bits generated per second between a pair of nodes.

% We denote the instantaneous SKR between a pair of nodes $i$ and $j$ at time time $t$ by $S_{ij}(t)$.

It is important to note that the entangled photon source has a limited capacity for distributing entangled photon pairs to various pairs of QKD nodes simultaneously. For instance, as illustrated in Fig.~\ref{fig:QCN}, the photon source is currently pumping entangled photon pairs to two  pairs of QKD nodes (indicated by the orange links).
% \textcolor{red}{why it should be distinct pairs, can a node be a member of two links? Are there any interference constraints?}. 
This is a key constraint on the photon source, as it directly influences the rate at which keys can be generated throughout the QN. Furthermore, as mentioned in the introduction, our analysis is based on the assumption that only a single optical link is used between any given pair of QKD nodes for photon distribution.
\subsection{State Descriptor for QN}
In this subsection, we define the graph structure representing our QN with a single photon source and describe the instantaneous SKR in this setting.
% We consider a QN with
% $n$ QKD nodes and a single polarization-entangled photon source as shown in Figure~\ref{fig:QCN}. 

\subsubsection{Graph Structure}The role of the photon source is to pump entangled photons between pairs of nodes. If the source provides entangled photons between a pair of nodes $(i,j)$, it establishes a connection between these two nodes. Therefore,
at each time $t$, the state of the QN can be represented by a graph  $G_t = (\mathcal{V}, \mathcal{E}_t)$, where
$
\mathcal{V} = \{1, 2, \dots, n\}
$
is the set of $n$ QKD nodes in the network, and
\[
\mathcal{E}_t = \{(i, j) : \text{\small{The source is pumping photons between }} i \text{\small{ and }} j \},
\]
is the set of edges present in the graph $G_t $. Each edge $(i, j) \in \mathcal{E}_t $ indicates that the source is actively pumping entangled photons to nodes $i$ and $j$ at time $t$. 
The graph $G_t $ reflects the flexibility of the photon source in pumping entangled photons to different pairs of QKD nodes over time. This provides a framework to analyze the connectivity and resource allocation in the QN.

We also define the feasible topology set $\mathcal{G} $, which comprises all $n$-node graphs $G = (\mathcal{V}, \mathcal{E})$.
For instance, the constraint that the photon source can simultaneously support at most $C$ connections results in the set
$\mathcal G = \{ ( \mathcal{V},\mathcal{E}) : |\mathcal{E}| \leq C \}$.  
 The set $\mathcal{G} $ is important in our framework because it defines the collection of network configurations that are both physically and practically viable for our QN, given constraints such as the limited capacity of the photon source.

\subsubsection{Instantaneous SKR}
At any time $t$, the SKR between nodes $i$ and $j$ depends on both the current network graph $G_t$ and the instantaneous channel conditions $\bm \chi(t)=(\chi_{ij}(t):i\in[n], j\in[n], i\leq j)$, where $\chi_{ij}(t)$ denotes the channel state between nodes $i$ and $j$. Therefore, we denote the instantaneous SKR between a pair $(i,j)$ at time $t$ as
\[
S_{ij}(G_t,\bm \chi(t)). %\,\indic{(i,j)\in \mathcal{E}_t },
\]
Note if a pair of nodes are not pumped, then their SKR is zero, i.e., if $(i,j) \notin \mathcal E_{t}$ then $S_{ij}(G_t,\bm \chi(t))=0$. 
Critically, the network topology $G_t$ is dynamically adjusted based on the observed channel state $\bm{\chi}(t)$. By continuously monitoring $\bm{\chi}(t)$, the network adapts its configuration in real time to optimize performance. 

We assume $\bm{\chi}(t)$ evolves as an independent and identically distributed (i.i.d.) process over time, with a stationary distribution $ \pi$. The channel state $\bm{\chi}(t) $ takes values from a finite set $ \mathcal{X}$ , where each $ \bm{\chi} \in \mathcal{X}$ represents a distinct configuration of channel conditions (e.g., high/low transmissivity, varying QBER). Here, $\pi$ is a probability distribution over $ \mathcal{X}$, such that $ \pi_{\bm{\chi}}$ denotes the steady-state probability of observing channel state $\bm{\chi}$ and 
$$
\sum_{\bm{\chi} \in \mathcal{X}}\pi_{\bm{\chi}}=1.
$$
While real-world channels may exhibit temporal correlations, the i.i.d. finite-state model provides a tractable foundation for analyzing fairness-performance trade-offs under stochastic dynamics. It is possible to extend the results below when the process $\bm \chi(t)$ evolves as a Markov chain, we refer the reader to \cite{stolyar2005asymptotic}.
% \textcolor{red}{Clarify how $G_t$ and $\chi(t)$ are related.}
% where the indicator function $\indic{(i,j)\in \mathcal{E}_t}$ ensures that a non-zero SKR is assigned only when there is an active connection between nodes $i$ and $j$. For notational simplicity, we drop the indicator and refer to it as $S_{ij}(G_t,\chi(t))$.

% For our theoretical analysis, we assume SKRs do not depend on instantaneous channel conditions that is $\bm{S}(G_t)=\{S_{ij}(G_t):i\in[n],j\in[n],i\leq j\}$.
% However, as we discuss, an extension to this case is possible. 
% Moreover, we numerically (Section~\ref{sec:evaluation}) test these proposed strategies under varying channel conditions. 
%Together, the graph $G_t$ and the SKR vector $\bm{S}(G_t)$ jointly characterize the dynamics of the QN considered in this work.

\subsection{SKR Rate Region}

The rate region $\mathcal{R}$ characterizes the set of all achievable average SKR between every pair of QKD nodes in the network. 

For every $\bm\chi \in \mathcal X$, let $\bm{P}_{\bm \chi} = (P_{G,\bm \chi} : G \in \mathcal{G})$ denote a probability distribution over graphs $\mathcal{G}$, where $P_{G,\bm \chi}$ is the probability that the network schedules graph $G$ when the channel state is $\bm \chi$.
Also let $\bm P = ( \bm{P}_{\bm \chi} : \bm \chi \in \mathcal X)$.
% It is defined by  the flexibility of the photon source to probabilistically schedule connections between nodes.
Then, we can write the average SKR vector $\bm{y}(\bm{P})$ as a convex combination of the instantaneous SKR vectors of all feasible topologies, weighted by their probabilities  
\[
\bm{y}(\bm{P}) = \sum_{\bm \chi \in \mathcal X} \pi_{\bm \chi}\sum_{G \in \mathcal{G}} P_{G,\bm \chi}  \bm{S}(G,\bm \chi).
\]  
Note that $\bm{y}(\bm{P})$ represents the long-term average SKR between every node pair where each $y_{ij}(\bm{P})$ is the expected SKR for the pair $(i,j)$ over time.
Therefore, we can define the rate region $\mathcal{R}$ as the set of all achievable average SKR vectors as
\begin{equation}\label{eq:R}
\mathcal{R} = \left\{ \bm{y}(\bm{P}) : P_{G,\bm\chi} \geq 0, \, \sum_{G \in \mathcal{G}} P_{G,\bm\chi} = 1, \forall \bm\chi \in \mathcal X \right\}. 
\end{equation}

\section{Problem statement and optimization formulation}
\label{sec:problem_statement}
% \neil{add intro sentence for structure here.}
In this section, we discuss the problem of pumping QKD nodes within our QN and formulate the corresponding optimization problem.

\subsection{Problem Statement}
The main challenge for the QN discussed above at any given time $t$ is determining how to efficiently pump entangled photons to QKD nodes so that they can generate secret keys. This requires deciding the graph $G_t $ given the instantaneous SKRs available currently and the average SKRs obtained in the past.
% , where $\mathcal{G} $ represents the collection of all feasible connection topologies between the $n$ users. 
The problem has two primary competing objectives:
\begin{enumerate}
    \item \textit{Fairness}: A fair algorithm must ensure that all pairs of nodes receive a stream of secret keys. 
    Thus, to ensure fairness in the selection of $G_t $, connections for nodes with poor channel conditions may need to be prioritized even if that results in a lower instantaneous SKR or throughput. Factors such as the physical distance between QKD nodes and environmental effects can degrade the quantum channel, resulting in a lower SKR.  This approach helps ensure that all the nodes can generate keys, rather than having only those with optimal conditions benefit repeatedly.
    \item \textit{Maximizing Throughput}: 
On the other hand, maximizing throughput focuses on enhancing the overall performance of the QN by selecting connections that yield the highest total SKR. This usually means favoring links with excellent channel conditions, where the photon source can deliver entangled photons more effectively. However, this approach risks repeatedly connecting the best-performing node pairs, potentially leaving nodes with poorer channels under-served.
    
    % The network should also aim to maximize the total SKR, which corresponds to the sum of the SKRs across all active connections in $G_t $. Achieving this requires choosing $G_t $ such that the total throughput is optimized while considering fairness constraints. The throughput for the QCN considered in this work depends on multiple factors such as: 1) The channel conditions between different user pairs. 2) The number of connections the photon source can simultaneously support. 3) The dynamic nature of quantum entanglement distribution.
    % \item \textit{Adapting to varying channel conditions}: \neil{Sanidhay please complete this...}
\end{enumerate} 

The tradeoff between fairness and throughput is central to communication network design. If we focus solely on throughput, we might neglect QKD nodes with bad channel conditions, leading to long-term inequities in key generation across QNs. Conversely, if we emphasize fairness too heavily, we might compromise the network's overall efficiency and reduce the total SKR. This tradeoff is not well understood in a QN setting, this can lead to systems that are jointly suboptimal with respect to throughput and fairness. Next, we provide an optimization formulation of network design that accounts for both throughput and fairness.
%Thus, the connection strategy for $G_t $must allow a balance, it should maximize the total throughput while ensuring that every pair of nodes, regardless of its channel condition, has enough opportunities to generate secret keys.

\subsection{Optimization Formulation}
% From the above discussion of balancing fairness and throughput in QN, we can formalize this trade-off using a utility-based optimization.
To balance fairness and throughput in a QN with time-varying channel conditions, we formalize this trade-off using a utility-based optimization framework. This approach accounts for stochastic channel dynamics, where the optimal pumping strategy must adapt to both instantaneous SKRs and historical performance.
This approach allows us to address the competing objectives of equitable and overall network performance in terms of total SKR.

% \neil{Maybe briefly provide referencing here.}

\subsubsection{Utility Function}
We define a utility function $U_{ij}(\bar{x}_{ij})$ for each pair of nodes $(i, j)$, where $\bar{x}_{ij}$ represents the average SKR allocated to the link between QKD nodes $i$ and $j$. The utility function $U_{ij}(\cdot)$ is chosen to be increasing and concave. 

Note that utility-based optimization is a well-established approach in classical networking. In traditional communication networks, similar utility functions have been used to balance fairness and throughput in resource allocation problems, as discussed in seminal works ~\cite{kelly1997charging,mo2000fair}. Our approach adapts these classical concepts to the quantum domain, ensuring that the optimization framework not only maximizes total SKR but also ensures equitable access to the limited photon source capacity.
% i)
% \textit{Increasing}: higher average SKRs correspond to greater utility, implying improved performance for the QKD node pair ii) \textit{Concave}: the gain in utility decreases as the average SKR increases, promoting a balanced resource allocation among QKD nodes and supporting fairness.

% \neil{get rid of i) and ii) probably not needed}

\subsubsection{Utility Optimization}
% The optimization problem can be formulated in terms of a utility function $U_{i,j}(S_{i,j}(t))$, where $U_{i,j}(\cdot)$ represents the utility function between users $(i, j)$ when a rate $S_{i,j}(t)$ is allocated. Assuming that $U_{i,j}(\cdot)$ is increasing and concave the objective is to maximize the sum of user utilities over a set of feasible topologies. 
% The main goal is to maximize the aggregate utility across all pair of QKD nodes in the QN.
% Therefore, at each time $t>0$ the goal is to select the graph $G_t $ such that:

The goal is to maximize the aggregate utility of the average SKRs across all node pairs in the QN. This requires optimizing over the long-term average SKRs $\bar{\bm{x}} = (\bar{x}_{ij})$ and probability distributions $\bm P = ( \bm{P}_{\bm \chi} : \bm \chi \in \mathcal X)$. Therefore, we have the following optimization problem
\begin{equation}
\label{eqn:opt}
\begin{aligned}
\text{Maximize} \quad & \sum_{i,j=1:i\leq j}^{n} U_{ij}(\bar{x}_{ij}) \\
\text{subject to} \quad & \bar{x}_{ij} = \sum_{\bm \chi \in \mathcal X} \pi_{\bm \chi}\sum_{G \in \mathcal{G}} P_{G,\bm \chi}  {S}_{ij}(G,\bm \chi), \quad \forall (i,j), \\
& P_{G,\bm \chi} \geq 0, \quad \forall G \in \mathcal{G}, \\
& \sum_{G \in \mathcal{G}} P_{G,\bm \chi} = 1, \quad \forall \bm \chi \in \mathcal{X},\\
\text{over}\quad &  \bm {\bar{x}} , \bm P \geq 0.
\end{aligned}    
\end{equation}
From the constraints of the above utility optimisation it is clear that the average SKR vector $\bar{\bm{x}}$ must lie within the rate region $\mathcal{R}$.
% , which is the convex hull of all feasible instantaneous SKR vectors $\{\bm{S}(G): G \in \mathcal{G}\}$. 

If $U_{ij}(\cdot)$ is concave, then this formulation results in a convex optimization problem. 
The choice of $U_{ij}(\cdot)$ determines the trade-off between fairness and throughput.
For example for proportional fairness we choose $U_{ij}(\bar{x}_{ij}) = \log(\bar{x}_{ij})$, then maximizing the sum of log is same as
\begin{equation}
\label{eqn:pf_objective}
 \text{Maximize} \quad \prod_{i,j=1:i\leq j}^n \bar{x}_{ij},   
\end{equation}
which is the product of average SKRs. This ensures no pair is starved, balancing fairness and throughput even under fluctuating $\bm{\chi}(t)$.
In the next section, we discuss a gradient-based algorithm that, when applied iteratively, converges to the solution of the optimization problem~\eqref{eqn:opt}.
% \begin{equation}
% \text{Maximize} \quad \sum_{(i,j) \in \mathcal{E}_t } U_{i,j}(x_{i,j}(t)), \quad \text{over} \quad G_t  \in \mathcal{G} ,
% \end{equation}

% subject to the following constraints:
% \begin{itemize}
%    %  \item \textit{Fairness Constraints}:  
%    % Ensure that users with poor channel conditions are also prioritized.
%    \item \textit{Network Constraints}:  The total number of edges in $ G_t $ is limited by the capacity of the photon source, denoted as $ C$:
% \begin{equation}
% |\mathcal{E}_t | \leq C.    
% \end{equation}
%      \item \textit{Positivity Constraint}: The average SKR for each connection must be non-negative:
% $$
% x_{i,j}(t) \geq 0, \quad \forall (i,j) \in \mathcal{E}_t .
% $$

%      \textit{Connectivity Constraints}:  
%    Ensure that each QKD nodes is connected to at least one other node over time to guarantee network-wide key generation that is
%  \begin{equation*}
%      \forall i \in \mathcal{V}, \quad \exists j \in \mathcal{V} \text{ such that } (i, j) \in \mathcal{E}_t .
% \end{equation*}

% \end{itemize}
% The above optimization ensures an efficient and fair allocation of entangled photon source capacity by dynamically selecting the graph $G_t $ that maximizes the total utility with different constraints.

% \neil{the optimization problem above is not well described as we don't relate $x$ to the choice of $G$. We also maybe don't want to include $t$ here either...}

\section{Gradient Based Pumping Algorithm}
\label{sec:gradient_algo}

We now discuss the gradient-based pumping algorithm (Algorithm~\ref{alg:gradient_pumping}) used to solve the utility maximization problem~\eqref{eqn:opt} by iteratively adjusting the average SKRs for each pair of QKD nodes in the QN. Similar algorithm for utility-maximization in classical switch is used by~\cite{stolyar2005asymptotic}.
% \begin{algorithm}[t!]
% \caption{Gradient-Based Pumping Algorithm}
% \label{alg:gradient_pumping}
% \begin{algorithmic}[1]
% \State \textbf{Initialize:} Set initial average SKRs $\bar{x}_{ij}(0)$ for all node pairs $(i,j)$.
% \For{$t = 1, 2, \dots$}
%     \State \textbf{Compute gradient weights:} For each pair $(i,j)$, calculate 
%     \[
%     w_{ij}(t) = U'_{ij}(\bar{x}_{ij}(t)).
%     \]
%     \State \textbf{Topology selection:} Determine the optimal topology
% \begin{equation}
% \label{eqn:optimal_top}
%     G^\star_t = \arg\max_{G \in \mathcal{G}} \sum_{i,j=1:i\leq j}^{n} w_{ij}(t) S_{ij}(G).    
% \end{equation}

%     \For{each node pair $(i,j)$}
%         \State \textbf{Update average SKR:} 
%         \begin{equation}
%         \label{eqn:update_rule}
%         \bar{x}_{ij}(t+1) = \bar{x}_{ij}(t) + \gamma(t) \bigl(S_{ij}(G^\star_t) - \bar{x}_{ij}(t)\bigr),
%         \end{equation}

%         where $\gamma(t)=\frac{1}{t+1}$ is a step-size parameter.
%     \EndFor
% \EndFor
% \end{algorithmic}
% \end{algorithm}
\begin{algorithm}[t!]
\caption{Gradient-Based Pumping Algorithm}
\label{alg:gradient_pumping}
\begin{algorithmic}[1]
\State \textbf{Initialize:} Set initial average SKRs $\bar{x}_{ij}(0)$ for all node pairs $(i,j)$.
\For{$t = 1, 2, \dots$}
    \State \textbf{Observe channel state:} Measure the current channel state $\bm \chi(t)$.
    \State \textbf{Compute gradient weights:} For each pair $(i,j)$, calculate 
    \[
    w_{ij}(t) = U'_{ij}(\bar{x}_{ij}(t)).
    \]
    \State \textbf{Topology selection:} Determine the optimal topology, given $\bm \chi(t)$, by solving
    \begin{equation}
    \label{eqn:optimal_top}
        G^\star_t = \arg\max_{G \in \mathcal{G}} \sum_{\substack{i,j=1 \\ i\leq j}}^{n} w_{ij}(t)\, S_{ij}(G,\bm \chi(t)).
    \end{equation}
    \For{each node pair $(i,j)$}
        \State \textbf{Update average SKR:}
        \begin{equation}
        \label{eqn:update_rule}
        \bar{x}_{ij}(t+1) = \bar{x}_{ij}(t) + \gamma(t) \bigl(S_{ij}(G^\star_t,\bm \chi(t)) - \bar{x}_{ij}(t)\bigr),
        \end{equation}
        where $\gamma(t)=\frac{1}{t+1}$ is a step-size parameter.
    \EndFor
\EndFor
\end{algorithmic}
\end{algorithm}

The algorithm operates in the following stages. At each time step $t$, the algorithm first observes the current channel state $\bm \chi(t)$, which influences the instantaneous SKRs $S_{ij}(G,\bm \chi(t))$. For each node pair, the derivative $U'_{ij}(\bar{x}_{ij}(t))$ of the utility function is computed to obtain the gradient weights $w_{ij}(t)$. These weights are then used to select the topology $G^\star_t$ that maximizes the weighted sum of the instantaneous SKRs, taking into account the current channel conditions. Once $G^\star_t$ is determined, the average SKR for each node pair is updated via a gradient ascent rule~\eqref{eqn:update_rule}, with the step-size $\gamma(t)=\frac{1}{t+1}$ controlling the convergence rate. This dynamic adjustment ensures that the algorithm adapts to time-varying channel conditions while balancing fairness and throughput.

\begin{rem}
The parameter $\gamma(t)$ in the update rule~\eqref{eqn:update_rule} controls the memory of the system. If $\gamma(t)$ is large, the instantaneous SKR values dominate. On the other hand if $\gamma(t)$ is smaller,
the average SKRs dominate, smoothing out fluctuations but slowing responsiveness.
When $\gamma(t) = 1/t+1 $, it implements a cumulative moving average. This ensures that $\bar{\bm x}(t)$
% $$ 
% \bar{x}_{ij}(t) = \frac{1}{t} \sum_{\tau=1}^t S_{ij}(G^\star_\tau),  
% $$ 
converges to the true average SKR over time. 
Therefore, the update gradually adjusts the long term average SKRs, ensuring that they converge toward the optimal operating point that maximizes the overall utility.
% The cumulative moving average guarantees asymptotic convergence to the optimal fair allocation.  
\end{rem}
Next we discuss different pumping strategies based on the specific choice of utility function in~\eqref{eqn:optimal_top}.

\section{Fairness and Pumping Strategies}
We first discuss the notion of fairness in relation to the utility function and explain different strategies for pumping entangled photons to different pair of nodes.
\label{sec:schemes}
\subsection{Fairness}
Fairness in QN is achieved by designing utility functions $U_{ij}(\bar{x}_{ij})$.
The $\alpha$-fairness utility framework provides a  way to model the trade-off between throughput and fairness. For an average SKR $\bar{x}_{ij}$ between nodes $i$ and $j$ the utility function is defined as:  
\begin{equation}
\label{eqn:utility}
 U_{ij}(\bar{x}_{ij}) = 
\begin{cases} 
\frac{\bar{x}_{ij}^{1-\alpha}}{1-\alpha}, & \alpha \neq 1, \\
\log(\bar{x}_{ij}), & \alpha = 1.
\end{cases}    
\end{equation}
Note that here $\alpha$ controls the fairness-efficiency balance. That is, 
$\alpha = 0$ maximizes total throughput (no fairness),  $\alpha = 1$ proportional fairness balances between throughput and fairness, and $\alpha \to \infty$ max-min fairness prioritizes the worst-off QKD node pair.

\subsection{Pumping Strategies}
The gradient-based algorithm adapts to time-varying channels by selecting topologies $G_t$ based on both the instantaneous channel state $\bm{\chi}(t)$ and historical SKRs $\bar{x}_{ij}(t)$. By selecting different utility functions $U_{ij}(\cdot)$, we can derive distinct pumping strategies that emphasize various trade-offs between fairness and throughput.
Specifically, we consider the following strategies, namely PF-PS, G-PS, and RR-PS. Below, we discuss each strategy in detail.
\subsubsection{PF-PS} 
This strategy balances fairness and throughput by prioritizing QKD node pairs based on their average SKRs and instantaneous SKR values adjusted for current channel conditions.
For this strategy we choose the logarithmic utility function $U_{ij}(\bar{x}_{ij}) = \log(\bar{x}_{ij})$. Then its derivative becomes $U'_{ij}(\bar{x}_{ij}) = 1/\bar{x}_{ij}$. In this case, the algorithm selects the topology at each time $t$ given $\bm{\chi}(t)$ as
\begin{equation}
\label{eQCN:pf_maximize}
    G^\star_t = G^\star(\bm {\bar x}(t), \bm \chi(t)) := \arg\max_{G \in \mathcal{G}} \sum_{\substack{i,j=1 \\ i\leq j}}^{n}  \frac{S_{ij}(G,\bm{\chi}(t))}{\bar{x}_{ij}(t)}. 
\end{equation}
PF-PS prioritizes links with high instantaneous SKR relative to their average SKRs. This scheme balances throughput and fairness as it penalizes over-allocated links with large $\bar{x}_{ij}(t)$.
A similar scheme has also been used for 4G wireless networks~\cite{viswanath2002opportunistic,andrews2001providing}.

\subsubsection{RR-PS}
As name suggest under this strategy all QKD node pairs receive equal chances to generate keys over time, irrespective of their channel conditions. We uses weighted  logarithmic utility function $
U_{ij}(\bar{x}_{ij}) = \frac{1}{S_{ij}}\log(\bar{x}_{ij}),$
whose derivative is $
U'_{ij}(\bar{x}_{ij}) = \frac{1}{S_{ij}\bar{x}_{ij}}$. Under the RR-PS at each time $t$ the topology is selected by maximizing the following objective
\begin{equation}
\label{eQCN:rr_maximize}
  G^\star_t = \arg\max_{G \in \mathcal{G}} \sum_{\substack{i,j=1 \\ i\leq j}}^{n}  \frac{1}{\bar{x}_{ij}(t)}. 
\end{equation} 
It is important to note that while fairness is prioritized under the RR scheme, this scheme does not maximize throughput, potentially leading to suboptimal resource utilization. Therefore, this scheme can also be seen as a maximum fairness achieving strategy.

\subsubsection{G-PS}
This strategy focuses on maximizing the SKR in the current time step based on channel condition $\bm \chi(t)$ without considering long-term fairness or average SKRs.
For the G-PS, we choose linear utility function $U_{ij}(\bar{x}_{ij}) = \bar{x}_{ij}$. Then the gradient becomes constant $U'_{ij}(\bar{x}_{ij})=1$, and the algorithm reduces to selecting the topology that maximizes the total instantaneous SKR that is
\begin{equation}
    G^\star_t = \arg\max_{G \in \mathcal{G}} \sum_{\substack{i,j=1 \\ i\leq j}}^{n}  S_{ij}(G,\bm \chi(t)). 
\end{equation}
% The graph $G_{t} $ is selected based on the current SKR values $S_{i,j}(t)$, selecting QKD node pairs with the highest values.

\section{Theoretical results for the PF-PS }
\label{sec:th_results}
Now we are in a position to prove the convergence result for the PF-PS scheme. We show that by iteratively applying the PF-PS scheme, the average SKR vector converges to the optimal solution of the optimization problem defined in~\eqref{eqn:opt} when the logarithmic utility function is employed.

\subsection{Differential Equation Formulation}

We discuss how the dynamics of a pumping strategy and, in particular, how the dynamics of the PF-PS can be represented by a set of differential equations. 
Notice from equation~\eqref{eqn:update_rule} that the discrete-time dynamics can be written as
% \textcolor{red}{why you have used $\alpha(t)$ instead of $\gamma(t)$ as in the algorithm?}
% \[
% \Delta x_{ij}(t) := x_{ij}(t+1) - x_{ij}(t) = \frac{1}{t_c} \left(
% S_{ij}(t) A_{ij}(t)  -x_{ij}(t)
% \right) 
% \]
\begin{equation}
\label{eqn:differn_eq}
 \Delta \bar{x}_{ij}(t) := \bar{x}_{ij}(t+1) - \bar{x}_{ij}(t) = \gamma(t) \left(
S_{ij}(G^\star_t, \bm \chi(t))  -\bar{x}_{ij}(t)
\right)   
\end{equation}
 
Notice that the average value of $S_{ij}(G^\star_t, \bm \chi(t)))$ in this update is 
\begin{align}
\bar S_{ij}(t) &:= \mathbb E_{\bm \chi} [S_{ij}(G^\star_t, \bm \chi(t))] \notag\\
&
=\sum_{\bm \chi \in \mathcal X} \pi_{\bm \chi}
    S_{ij} ( G^\star ( \bm {\bar{x}}(t), \bm \chi ) , \bm \chi)  \,.    \label{eq:Sbar}
\end{align}
Thus for small but fixed values of $\gamma(t)$, the recursion given in~\eqref{eqn:differn_eq} is a finite-difference approximation for the differential equation
\[
\frac{d\bar{x}_{ij}}{dt} = \gamma ( \bar S_{ij}(t) - \bar{x}_{ij}(t)) \, ,
\]
here we note that if we employ the PF-PS scheme, then $G^\star_t$ is obtained from~\eqref{eQCN:pf_maximize}.

% $A_{ij}(t)$ must solve the optimization
% \[
% \text{maximize}_{A}\quad \sum_{i,j} \frac{S_{ij}(t)A_{ij}}{x_{ij}(t)}
% \quad\text{ over }\quad A 
% \]

\subsection{Convergence of PF-PS}

Building on the previous discussion, we can now assert that the PF-PS update rule evolves in such a way that it converges toward the optimal average secret key rate, where the average is defined as the geometric mean of the individual key rates. 

\begin{theorem}
\label{thm:pf_sol} 
Assume that the average SKR for each user pair $(i,j)$, denoted by $\bar{x}_{ij}(t)$, evolves according to
\begin{equation}
\label{eqn:x_drt}
\frac{d\bar{x}_{ij}}{dt} = \gamma\Bigl(\bar S_{ij}(t) - \bar{x}_{ij}(t)\Bigr). %\quad \text{with} \quad \gamma(t)=\frac{1}{t+1}.
\end{equation}
% where $S_{ij}(G_t^\star)$ is the instantaneous SKR achieved when the photon source targets the user pair $(i,j)$ in the selected graph $G_t^\star$. 
Then, by iteratively selecting $G_t^\star$ using the PF-PS to maximize the objective in~\eqref{eQCN:pf_maximize}, the resulting sequence of average SKR vectors converge to the solution of maximization problem
\begin{equation}
\label{eqn:log_utility1}    
\begin{aligned}
\text{Maximize} \quad & \prod_{i,j=1:i\leq j}^{n} \bar{x}_{ij}^{\frac{1}{n}} \\
\text{subject to} \quad & \bar{x}_{ij} = \sum_{\bm \chi \in \mathcal X} \pi_{\bm \chi}\sum_{G \in \mathcal{G}} P_{G,\bm \chi}  {S}_{ij}(G,\bm \chi), \quad \forall (i,j), \\
%& P_{G,\bm \chi} \geq 0, \quad \forall G \in \mathcal{G}, \\
& \sum_{G \in \mathcal{G}} P_{G,\bm \chi} = 1, \quad \forall \bm \chi \in \mathcal{X}\\
\text{over }\quad & \bm {\bar{x}} , \bm P \geq 0
\end{aligned}    
\end{equation}

% % At each time $t$, selecting the graph $G_t^\star $ using the PF-PS, which maximizes the objective in~\eqref{eQCN:pf_maximize}. 
% If the average SKRs $ x_{ij}(t)$ for each user pair $ (i, j)$ evolves according to 
% \begin{equation}
% \label{eqn:x_drt}
% \frac{dx_{ij}}{dt} =  \alpha(t)(S_{ij}(G^\star_t) - x_{ij}(t)), \  \text{for }, \ \alpha(t)=1/(t+1).
% \end{equation}
% Then selecting the graph $G_t^\star $ using the PF-PS iteratively which maximizes the objective in~\eqref{eQCN:pf_maximize} in turns maximizes the logarithmic utility function that is \begin{equation}
% \label{eqn:log_utility}
% \max \sum_{i,j=1:i\leq j}^{n} \log(x_{ij}).    
% \end{equation}
% \neil{what is the sum over here.}

% At each time $t$, selecting the graph $ G_t $ using the PF scheme maximizes the following 
% \begin{equation}
% \label{eQCN:log_utility}
% \max_{G_t  \in \mathcal{G} } \sum_{(i,j) \in \mathcal{E}_t } \log(x_{i,j}(t)).    
% \end{equation}
% % Therefore, for every $t>0$ we have 

% % \[
% % G_{t,\text{PF}}  \in \arg\max_{G_t  \in \mathcal{G} } \sum_{(i,j) \in \mathcal{E}_t } \log(x_{i,j}(t)).
% % \]
\end{theorem}
Theorem~\ref{thm:pf_sol} establishes a key property of the PF-PS at any given time $t > 0$, the graph $G_t^\star$ selected according to the PF-PS maximizes the geometric mean of the SKR values for the QKD node pairs. In contrast, the G-PS is designed solely to maximize total throughput by selecting the graph that yields the highest instantaneous SKR sum, without considering historical averages or fairness. On the other hand, the RR-PS enforces strict fairness by giving all QKD node pairs equal opportunities to connect, irrespective of their current SKR values. Thus, the PF-PS is particularly beneficial for quantum networks with QKD nodes, as it dynamically adapts resource allocation to changing network conditions, effectively balancing throughput and fairness while mitigating the shortcomings of purely greedy or round-robin approaches.

{\em Proof of Theorem~\ref{thm:pf_sol}:} 
To prove the result we first collect together some bounds on our processes $\bm {\bar{x}}(t)$ and $\bm {\bar S}(t)$. This includes writing our objective as a utility maximization, then arguing the feasibility and sub-optimality of $\bm {\bar x}(t)$ in \eqref{eq:xsub}, below. The sub-optimality of $\bm x^\star$ for the PF-PS objective in \eqref{eq:SbarInequ}. 
Following this we combine these terms to prove the result.
%To prove that choosing $G_{t}^\star$ from~\eqref{eQCN:pf_maximize} iteratively maximizes~\eqref{eqn:log_utility}, we use~\eqref{eqn:x_drt}. %leads to the maximization of~\eqref{eqn:log_utility}.

Since the logarithm function is strictly increasing, it is immediate that maximizing the geometric mean in~\eqref{eqn:log_utility1} is equivalent to maximizing the sum of logarithms.
\begin{equation}
\label{eqn:log_utility}    
\begin{aligned}
\text{Maximize} \quad & \sum_{i,j=1:i\leq j}^{n} \log(\bar{x}_{ij}) \\
\text{subject to} \quad & \bar{x}_{ij} = \sum_{\bm \chi \in \mathcal X} \pi_{\bm \chi}\sum_{G \in \mathcal{G}} P_{G,\bm \chi}  {S}_{ij}(G,\bm \chi), \quad \forall (i,j), \\
%& P_{G,\bm \chi} \geq 0, \quad \forall G \in \mathcal{G}, \\
& \sum_{G \in \mathcal{G}} P_{G,\bm \chi} = 1, \quad \forall \bm \chi \in \mathcal{X}\\
\text{over }\quad & \bm {\bar x} , \bm P \geq 0
\end{aligned}    
\end{equation}

Let $(\bm x^*, \bm P^*)$ denote an optimal value at which~\eqref{eqn:log_utility} attains its maximum. 
% Thus, we see that
% \[
% \log(x_{ij}^\star) = \sum_{\substack{i,j=1 \\ i\leq j}}^{n} \log(v_{ij}^*).
% \]
Notice we can write $\bm {\bar S}$ from \eqref{eq:Sbar} as follows 
\begin{align*}
\bar S_{ij}(t) 
=\sum_{\bm \chi \in \mathcal X} \pi_{\bm \chi}
\sum_{G \in \mathcal G} I_{G,\bm \chi}(t)
    S_{ij} ( G, \bm \chi)    
\end{align*}
where
\[
I_{G,\bm \chi}(t) = 
\begin{cases}
    1 & \text{if } G = G^\star ( \bm {\bar x}(t), \bm \chi )\, ,\\
    0 & \text{otherwise.}
\end{cases}
\]
In other words, we see that $(\bm {\bar S}(t), \bm I(t))$, where $\bm I(t)=(I_{G,\bm \chi}(t): \bm \chi \in \mathcal{X})$ is the probability distribution, also a feasible solution (albeit sub-optimal) for the optimization \eqref{eqn:log_utility}. Further, since $\bm {\bar x}(t)$ is a convex combination of these terms it is also a feasible solution for the optimization \eqref{eqn:log_utility}. That is $\bm {\bar x}(t) \in \mathcal R $, as defined in \eqref{eq:R} and 
\begin{equation}\label{eq:xsub}
\sum_{i,j: i\neq j} \log x_{ij}^\star \geq \sum_{i,j: i\neq j} \log \bar x_{ij}(t).    
\end{equation}
While we see $\bm x^\star$ is optimal for the above logarithmic objective, $\bm {\bar S}(t)$ is superior for the objective of the PF-PS problem. In particular, note that from the definition of $\bm {\bar S}$ in \eqref{eq:Sbar}, and the definition of the PF-PS policy \eqref{eQCN:pf_maximize} we have
\begin{align}
&\sum_{i,j: i\neq j}  \frac{\bar{S}_{ij}(t)}{\bar x_{ij}(t)} \notag
=
\sum_{\bm \chi \in \mathcal X} \pi_{\bm \chi} 
\max_{G \in \mathcal{G}} \sum_{i,j: i\neq j} \frac{S_{ij}(G,\bm \chi)}{\bar x_{ij}(t)}
\notag \\&
\geq 
\sum_{\bm \chi \in \mathcal X} \pi_{\bm \chi} 
\sum_{G \in \mathcal{G}} P^\star_{G, \bm \chi} \sum_{i,j: i\neq j} \frac{S_{ij}(G,\bm \chi)}{\bar x_{ij}(t)}
=
\sum_{i,j: i\neq j} 
\frac{x_{ij}^\star}{\bar x_{ij}(t)}\, . \label{eq:SbarInequ}
\end{align}

We make a final observation about $\bm x^\star$ before proceeding with the proof. Note that for any other feasible $\bm x$, since the objective function is concave, it must hold that the directional derivative from $\bm x$ to $\bm x^\star$ must be positive and greater than the change in the function values. That is,
\begin{align}
& \frac{d \sum_{ij} \log(x_{ij})}{d \overset{\rightarrow}{\bm x\bm x^\star}} 
\notag
\\
&
:=
\lim_{h \searrow 0} \frac{1}{h} \left( \sum_{ij} \log(x_{ij}(t) + h( x_{ij}^\star - x_{ij}(t)))
- \log(x_{ij}(t))
\right) 
\notag
\\
&=  \sum_{ij} (x_{ij}^\star  - x_{ij} ) \frac{1}{x_{ij}} 
 \geq 
 \sum_{ij} \log (x_{ij}^\star) - \sum_{ij} \log (x_{ij}) > 0
 \,, \label{eq:direction}
\end{align}
for $\bm x \neq \bm x^\star$. The first inequality above holds by the concavity of the objective. The second follows by the sub-optimality of $x_{ij}$.
% From~\eqref{eQCN:avg_skr_update}, the time derivative of $x_{i,j}(t)$ for any $(i,j) \in \mathcal{E}_t $ can be written as:
% \begin{equation}
% \label{eQCN:x_drt}
% \frac{dx_{i,j}(t)}{dt} = \alpha (S_{i,j}(t) - x_{i,j}(t)).
% \end{equation}

Now, we compute the derivative of the logarithmic utility sum in~\eqref{eqn:log_utility} with respect to $t$:
\begin{align}
& \frac{d}{dt}\sum_{ij} \log(\bar x_{ij}(t)) \notag 
\\
 &=\sum_{ij} \frac{d \bar x_{ij}(t)}{dt} \times \frac{d\log(\bar x_{ij}(t))}{d\bar{x}_{ij}(t)} \nonumber\\
 &=\sum_{ij }\gamma (\bar S_{ij}(t)-\bar x_{ij}(t)) \times \frac{1}{\bar x_{ij}(t)}\nonumber\\
&\geq \sum_{ij}\gamma  x_{ij}^\star \times \frac{1}{\bar x_{ij}(t)} - \gamma \sum_{ij }\bar{x}_{ij}(t) \times \frac{1}{\bar{x}_{ij}(t)} \nonumber\\
 &\geq \sum_{ij}\gamma \log ( x_{ij}^\star)-\gamma \sum_{ij }\log (\bar x_{ij}(t))>0\nonumber,
\end{align}
% the second equality follows from~\eqref{eqn:x_drt}, the first inequality follows from 
% \[
% \max_{G_t  \in \mathcal{G} } \sum_{(i,j) \in \mathcal{E}_t } \frac{S_{i,j}(t)}{x_{i,j}(t)}
% \geq \sum_{(i,j) \in \mathcal{E}_{t,*} } \frac{x_{i,j}^\star}{x_{i,j}(t)},
% \]
To explain the calculations above: The first equality applies the chain rule. Then we apply the definition of the differential equation \eqref{eqn:x_drt}. Then we apply inequality \eqref{eq:SbarInequ}. And finally, we apply \eqref{eq:direction}. Since the set of values of $\mathcal R$ is a closed bounded convex set, the value of $\bm {\bar x}(t)$ must converge to $\bm x^\star$. 
$\Box$

\begin{rem}
There are several possible extensions to the model presented above. First, by adopting a different utility function, the framework can be extended to optimize alternative objectives. Second, although our current analysis assumes i.i.d. noise, the same approach can be adapted to handle Markovian time-varying channel conditions. Finally, the model could be further refined to incorporate terms that account for the number of secure keys shared between nodes. We leave these extensions as directions for future work.
\end{rem}

\section{Example of QN with $n=4$ QKD nodes}
\label{sec:numerical_studies}

We now explain the working of each strategy with an illustrative example. To do so, we consider a QN with four  QKD nodes ($n = 4$) and a photon source with a capacity $C = 2$. This capacity constraint implies that the photon source can support at most two simultaneous edges (connections) between user pairs at any given time. Consequently, the set of all possible topologies is denoted by $\mathcal G = \{ ( V,E) : |E| \leq 2 \}$.

In this example our main aim is to observe how each scheme PF-PS, RR-PS, and G-PS selects a graph $G_t^\star \in \mathcal{G}$ at each time step. 
% The fixed SKR matrix $\bm{S}$ and the initial average SKR vector $\bar{\bm{x}}(0)$ are given by
% \[
% \bm{S} = \begin{bmatrix}
% 0 & 100 & 200 & 300 \\
% 100 & 0 & 400 & 500 \\
% 200 & 400 & 0 & 600 \\
% 300 & 500 & 600 & 0
% \end{bmatrix}, \quad
% \bar{\bm{x}}(0) = \begin{bmatrix}
% 0 & 10 & 20 & 30 \\
% 10 & 0 & 20 & 50 \\
% 20 & 20 & 0 & 60 \\
% 30 & 50 & 60 & 0
% \end{bmatrix}.
% \]
For simplicity, we assume fixed channel conditions (i.e., the instantaneous SKR matrix $\bm{S}$ does not vary over time). This allows us to isolate the impact of each strategy’s resource allocation logic.
Therefore, we consider the following SKR matrix $\bm S$
\begin{equation*}
\bm S= 
\begin{bmatrix}
0 & 100 & 200 & 300 \\
100 & 0 & 400 & 500 \\
200 & 400 & 0 & 600 \\
300 & 500 & 600 & 0
\end{bmatrix},    
\end{equation*}
where $S_{ij}$ represents the SKR for edge $(i, j)$, and $S_{ij}= S_{ji}$ (symmetric matrix). The highest SKRs are observed for pairs $(3,4)$, $(2,4)$, and $(2,3)$, with values $600$, $500$, and $400$, respectively.
% The average SKR $\bar{\bm x}$ is initialized as $\bar{\bm x}=\bm S$.
The initial average SKRs are set as
$\bar{x}_{i,j}(0) = 10, \quad \forall\, (i,j),$ with $\bar{x}_{i,j}=\bar{x}_{j,i}$ and employs an moving average with parameter $\gamma=0.5$.

PF-PS balances fairness and throughput by dynamically adjusting weights based on historical allocations. Initially ($t=1$), it selects $(3,4)$ and $(2,4)$, as their high SKRs dominate the unweighted selection. However, by $t=2$, the updated averages $\bar{x}_{3,4}(1) = 305$ and $\bar{x}_{2,4}(1) = 255$ reduce their weights ($w_{i,j} = S_{i,j}/\bar{x}_{i,j} $), prompting PF-PS to shift focus to the next-highest weighted pair, $(2,3)$, which now offers a higher marginal utility. This adaptability ensures that under-allocated pairs gradually receive attention, balancing long-term fairness without sacrificing significant throughput.

% Now, for the PF-PS, we select $G^\star_{t} $, the topology that achieves the maximum in~\eqref{eQCN:pf_maximize}. At $t=0$, the topology $G^\star_{0}  = \{(2, 3), (3, 4)\}$ achieves the highest proportional fairness objective, as it balances high SKRs with lower average SKRs.

% The RR-PS selects the topology that maximizes~\eqref{eQCN:rr_maximize}. In this case, edges with lower average SKRs are prioritized. At $t=0$, the topology $G^\star_{0}  = \{(1, 2), (2, 3)\}$ is selected.
RR-PS enforces strict fairness by cycling through all pairs. At $t=1$, it prioritizes the least-served pairs $(1,4)$ and $(1,3)$, updating their averages to $\bar{x}_{1,4}(1)=155$ and $\bar{x}_{1,3}(1)=105$. By $t=2$, it rotates to $(1,2)$ and $(2,3)$, raising their averages to $\bar{x}_{1,2}(2)=55$ and $\bar{x}_{2,3}(2)=205$. While RR-PS ensures equitable resource distribution, it achieves the lowest total SKR (e.g., $155 + 105 = 260$ at $t=1 $) compared to G-PS and PF-PS.

G-PS prioritizes immediate throughput by selecting the two highest-SKR pairs at each time step. In the first iteration ($t=1$), it connects $(3,4)$ and $(2,4)$, achieving a total SKR of $600 + 500 = 1100$. By the second iteration ($t=2$), it continues selecting the same pairs, further increasing their average SKRs to $\bar{x}_{3,4}(2)=455$ and $\bar{x}_{2,4}(2)=377.5$, respectively. While G-PS maximizes total throughput, it risks resource monopolization, as low-SKR pairs like $(1,2)$ and $(1,3)$ remain neglected, decaying to $\bar{x}_{i,j} = 10$.

\section{Numerical Comparison}
\label{sec:evaluation}
For numerical studies we have chosen a QN comprising of $n=5$ QKD nodes. The network topology is represented by two key matrices one for the physical distances between the nodes and another for the corresponding QBER for each link.

The distances between each pair of nodes are defined in a symmetric matrix $\bm L=(L_{ij})$, where $L_{ij}$ represents the physical separation (in kilometers) between node $i$ and $j$. 
% For the example, QKD nodes $1$ and $2$ are $10$ km apart, nodes $1$ and $3$ are $20$ km apart. 
This structured distance distribution allows us to simulate realistic link losses in a QN. We set $\bm L$ as shown in Table~\ref{tab:distance_mat}.

% \begin{table}[h]
% \centering
% \begin{tabular}{c|ccccc}
% \textbf{}     & \textbf{Node 1} & \textbf{Node 2} & \textbf{Node 3} & \textbf{Node 4} & \textbf{Node 5} \\ \hline
% \textbf{Node 1} & 0    & 800  & 20  & 7   & 400 \\
% \textbf{Node 2} & 800  & 0    & 150 & 25  & 35  \\
% \textbf{Node 3} & 20   & 150  & 0   & 10  & 20  \\
% \textbf{Node 4} & 7    & 25   & 10  & 0   & 150 \\
% \textbf{Node 5} & 400  & 35   & 20  & 150 & 0   \\
% \end{tabular}
% \caption{Distance matrix between QKD nodes}
% \label{tab:distance_mat}
% \end{table}

\begin{table}[h]
\centering
\begin{tabular}{c|ccccc}
& \textbf{Node 1} & \textbf{Node 2} & \textbf{Node 3} & \textbf{Node 4} & \textbf{Node 5} \\ \hline
\textbf{Node 1} & 0 & 50 & 80 & 20 & 100 \\
\textbf{Node 2} & 50 & 0 & 30 & 60 & 90 \\
\textbf{Node 3} & 80 & 30 & 0 & 70 & 40 \\
\textbf{Node 4} & 20 & 60 & 70 & 0 & 10 \\
\textbf{Node 5} & 100 & 90 & 40 & 10 & 0 \\
\end{tabular}
\caption{Distance matrix between QKD nodes}
\label{tab:distance_mat}
\end{table}

Alongside the distance matrix, a QBER matrix $\bm {Qb}$ (Table~\ref{tab:qber_matrix}) is defined where each off-diagonal element indicates the error rate on the corresponding link.
Moreover, we have chosen the fiber attenuation coefficient $\beta=0.2$ dB/km.
% \begin{table}[h]
% \centering
% \begin{tabular}{c|ccccc}
% \textbf{}     & \textbf{Node 1} & \textbf{Node 2} & \textbf{Node 3} & \textbf{Node 4} & \textbf{Node 5} \\ \hline
% \textbf{Node 1} & 0       & 0.1   & 0.03  & 0.001 & 0.08 \\
% \textbf{Node 2} & 0.1     & 0     & 0.05  & 0.03  & 0.04 \\
% \textbf{Node 3} & 0.03    & 0.05  & 0     & 0.02  & 0.03 \\
% \textbf{Node 4} & 0.001   & 0.03  & 0.02  & 0     & 0.01 \\
% \textbf{Node 5} & 0.08    & 0.04  & 0.03  & 0.01  & 0    \\
% \end{tabular}
% \caption{QBER matrix between QKD nodes channels}
% \label{tab:qber_matrix}
% \end{table}

\begin{table}[h]
\centering
\begin{tabular}{c|ccccc}
& \textbf{Node 1} & \textbf{Node 2} & \textbf{Node 3} & \textbf{Node 4} & \textbf{Node 5} \\ \hline
\textbf{Node 1} & 0 & 0.02 & 0.03 & 0.005 & 0.04 \\
\textbf{Node 2} & 0.02 & 0 & 0.015 & 0.025 & 0.035 \\
\textbf{Node 3} & 0.03 & 0.015 & 0 & 0.03 & 0.02 \\
\textbf{Node 4} & 0.005 & 0.025 & 0.03 & 0 & 0.005 \\
\textbf{Node 5} & 0.04 & 0.035 & 0.02 & 0.005 & 0 \\
\end{tabular}
\caption{QBER matrix between QKD nodes channels}
\label{tab:qber_matrix}
\end{table}
\begin{rem}
  We considered a $5$ node QKD network simply to demonstrate and verify our proposed fairness algorithms. 
  From a QKD point of view, every link here is at most 100 km within the range where direct fiber QKD works without any repeaters. However, the network still highlights the key challenges of edge selection, fairness, and error‑rate impacts on secret key generation.
   Crucially, our formulation and centralized controller for pumping entangled photons extend directly to any local‐area QKD network of size $n$. 
\end{rem}

At time $t$, the SKR $S_{ij}(G^\star_t)$ between node $i$ and node $j$ is computed based raw key rate  
% $$
% S_{ij}(G^\star_t)= R_{ij}  \left[1 - h(Qb_{ij}) \right],
% $$ 
$R_{ij}  =\eta_{ij} \cdot r$ with $r$ being the repetition rate, and the binary entropy $h(Qb_{ij})$. Specifically, $
 S_{ij} \approx R_{ij}  \Big( 1 - h\big(Qb_{ij}\big) \Big),$
where $h(x)=-x\log x -(1-x) \log(1-x)$ is the binary entropy function. 
% (see Eq 43 in~\cite{scarani2009security}) 
Furthermore, the QBER increases with the distance $L_{ij}$ between nodes  $i$ and $j$. This relationship arises because $
   \eta_{ij} = 10^{-\alpha L_{ij}/10}$~\cite{scarani2009security}.

In our numerical studies, we consider two distinct scenarios to evaluate the performance of various key generation strategies across a QN. These scenarios are designed to capture both static and dynamic behaviors of the network’s SKR across the links between nodes.
\begin{itemize}
    \item  \textit{Fixed instantaneous SKR across QKD nodes}: This is a valid assumption when the channel conditions such as fiber attenuation, and  environmental noise exhibit minimal variation over the time. 
    
    % By assuming a fixed SKR, we establish a baseline performance for the QN, allowing us to analyze the capabilities of the QKD system under quasi-static conditions.

    \item \textit{Time-varying SKR (changing every $T$ units of time)}: In this case channel conditions may vary due to like atmospheric disturbances, or fiber aging, leading to time dependent variations in both transmissivity and QBER. By incorporating time varying SKR, we can evaluate the adaptability of different key generation strategies.
\end{itemize}

\subsection{Fixed instantaneous SKR across QKD nodes}
\begin{figure}[h!]
\centering
  \medskip
  \centering
  \includegraphics[width=7.5cm]{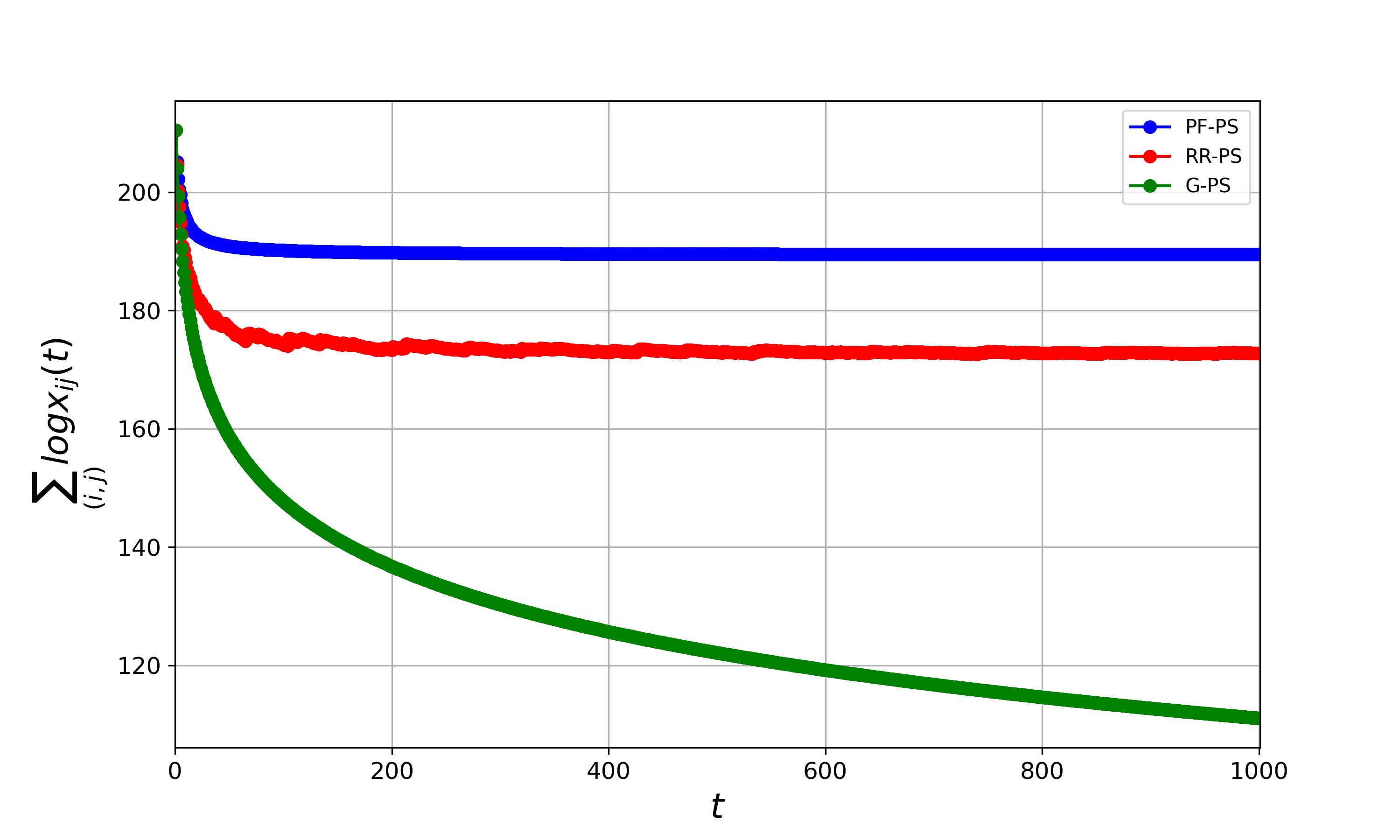}
\caption{Performance comparison for all three strategies for QN with $n=5$ QKD nodes.}
\label{fig:scheme_comp}
\end{figure}

To compare the performance of each strategy under this scenario, Figure~\ref{fig:scheme_comp} illustrates the logarithmic sum of the average SKR values for QKD pairs, $\sum_{(i,j) } \log(\bar{x}_{ij}(t))$, as a function of time $t$.
From the figure, it is evident that the PF-PS achieves the highest logarithmic sum over time. This indicates that PF-PS balances the SKR allocation across QKD pairs, ensuring fairness and preventing starvation of less-utilized edges.
In contrast, G-PS selects the same pair of QKD nodes for pumping photons for all time slots, focusing only on maximizing the immediate SKR sum. Consequently, only the selected QKD pairs receive updates to their average SKR values $\bar{x}_{ij}$, while the remaining user pairs are left idle, unable to generate keys. Over time, this behavior causes the SKRs for non-selected pairs to diminish, effectively leading to starvation. This phenomenon is reflected in the plot, where the logarithmic sum $\sum_{(i,j)} \log(\bar{x}_{ij}(t))$ steadily declines and eventually approaches zero as time progresses.
RR-PF lies in between, as it prioritizes edges with lower historical averages, providing a more equitable distribution than G-PS but still lagging behind PF-PS.

\subsection{Time-varying SKR (changing every $T$ units of time)}
For our simulations, we set $T = 100$, meaning that every $100$ time slots the SKR matrix is updated to reflect new channel conditions. Specifically, we modify the QBER values across different links, which in turn alters the achievable SKR on those links.
This is achieved through a controlled perturbation process that introduces small, randomized fluctuations to the QBER values of existing links while preserving physical constraints. 
For each link $(i, j)$ in the QBER matrix a uniformly distributed random variation $\delta \in [-0.005, +0.005]$ is added to the current QBER value. This simulates minor fluctuations in channel noise and loss caused by environmental factors such as temperature changes, fiber stress, or component drift. The perturbed QBER is clipped to the range $[0.0, 0.5]$ to ensure values remain physically meaningful. A QBER of $0.5$ represents the error rate threshold beyond which secure key distribution becomes impossible. 
\begin{figure}[h!]
\centering
  \medskip
  \centering
  \includegraphics[width=7.5cm]{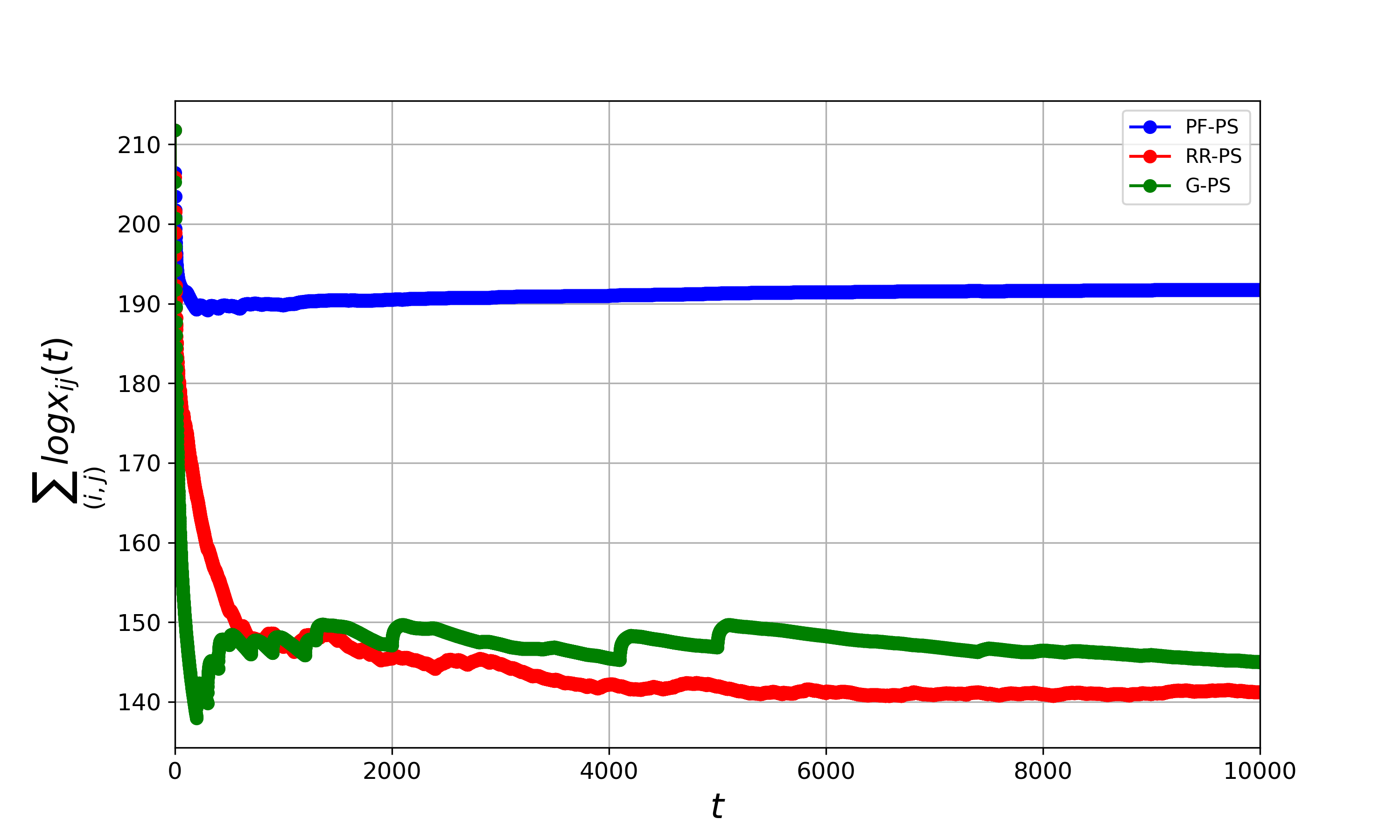}
\caption{Time-varying channel after every $T=100$.}
\label{fig:variable_skr}
\end{figure}

% \textcolor{red}{What are the used channel conditions and qber values?}
In this dynamic scenario, PF-PS consistently outperforms both G-PS and RR-PS. The key advantage of PF-PS is its ability to reassign resources to improving links while preserving long-term fairness through historical averages. Meanwhile, RR-PS performs the worst under time-varying SKR because it maximizes the objective in \eqref{eQCN:rr_maximize} based solely on inverse historical averages, thus failing to adapt when link conditions change. Although G-PS does respond to the current SKR by selecting the best links at each time step, it still neglects fairness and historical usage, making it less useful than PF-PS in maintaining performance over the long run.

\section{Conclusion and Future Directions}
\label{sec:conclusion}

% We addressed the problem of optimizing pumping strategies in QNs designed for secure communication via QKD. Our approach is broadly applicable to any bipartite entanglement-based QKD protocols like BBM92. The QN is modeled as a time-varying graph, where edges represent active links between user pairs that distribute entangled photons. We formulated an optimization problem aimed at selecting the best connections to pump entangled photons at each time slot to maximize a utility function. We evaluated three distinct pumping strategies PF-PS, G-PS, and RR-PS. Our analysis, supported by numerical simulations, demonstrates that PF-PS, which dynamically prioritizes links with low historical average SKRs while leveraging instantaneous SKR, achieves a superior balance between fairness and throughput compared to the other strategies.
In this work, we addressed the problem of optimizing entanglement distribution in QN with QKD nodes under time-varying channel conditions.  
% Our framework, applicable to bipartite entanglement-based protocols like BBM92.
We models the QN as a time-varying graph, with edges representing active links dynamically adjusted to adapt to stochastic channel states. Furthermore, we formulated an optimization problem aimed at selecting the best connections to pump entangled photons at each time slot to maximize a utility function. To solve this optimization problem we introduced a gradient driven pumping algorithm that balance the fairness and throughput under these variable conditions.  
The proposed PF-PS dynamically prioritizes links with high instantaneous SKRs relative to their historical averages, effectively balancing fairness and throughput even as channel conditions changes. 
% This ensures equitable resource allocation without sacrificing throughput, making PF-PS robust to varying channel conditions. 
Numerical simulations for both fixed and time-varying QBER regimes demonstrate PF-PS’s superiority. While G-PS maximizes short-term throughput, it leads to inequities under channel fluctuations. Conversely, RR-PS enforces strict fairness but suffers from suboptimal throughput as it ignores real-time channel state. 
To our knowledge, this is the first work to apply a gradient-based algorithm for utility optimization in QNs. 
% This method enables real-time adaptation to channel dynamics, a critical advantage in practical, unpredictable environments.  

Our immediate future directions involve extending the current single-hop model to more complex, multi-hop networks, where the SKR between any two nodes is influenced by the performance of intermediate nodes. 
% This raises important questions about the adaptability of the proposed pumping strategies in multi-hop scenarios.
Additionally, we plan to incorporate the demand (number of keys) dynamics of individual QKD nodes into our framework. Specifically, it will be interesting to explore the scenario where nodes with abundant keys can support higher communication rates, and mechanisms will be developed to ensure that nodes do not deplete their key reserves below critical levels. 
% These advancements will contribute to the design of robust, efficient, and adaptable QKD systems in dynamic environment.

\textbf{Acknowledgements:}
 This research was funded by the EPSRC funded INFORMED-AI project EP/Y028732/1. Siddarth Koduru Joshi would like to acknowledge funding from UK Research and Innovation's Engineering and Physical Science Research Council for the Integrated Quantum Networks hub and the new investigator grant Towards the quantum internet, EP/Z533208/1 and EP/X039439/1, respectively.

\bibliographystyle{IEEEtran}
\bibliography{ref} 
\end{document}